# Crossover critical behavior of $Ga_{1-x}Mn_xAs$

Sh. U. Yuldashev,[1,*] Kh. T. Igamberdiev,[1], Y. H. Kwon,[1] Sanghoon Lee,[2] X. Liu,[3] J. K. Furdyna,[3] A. G. Shashkov,[4] and T. W. Kang[1]

[1]*Quantum-Functional Semiconductor Research Center, Dongguk University, Seoul 100-715, Korea*

[2]*Department of Physics, Korea University, Seoul 136-701, Korea*

[3] *Department of Physics, University of Notre Dame, Notre Dame, Indiana 46556, USA*

[4]*Institute of Heat and Mass Transfer, 15P. Brouki, Minsk 220072, Belarus*

The critical behavior of $Ga_{1-x}Mn_xAs$ in a close vicinity of the Curie temperature was experimentally studied by using the thermal diffusivity measurements. Taking into account that the inverse of the thermal diffusivity has the same critical behavior as the specific heat, the critical exponent α for the samples investigated has been determined. With approaching close to the critical temperature, the crossover from the mean-field-like to the Ising-like critical behavior has been observed. From the crossover behavior the values of the Ginzburg number and the exchange interaction length in $Ga_{1-x}Mn_xAs$ with different concentrations of Mn were determined.

[*]E-mail address: shavkat@dongguk.edu

The critical behavior in the proximity of the Curie temperature is still one of the central problems in the physics of itinerant ferromagnets. By establishing the universality class for the phase transition, information is provided on the range of the exchange interactions. Examples are, a long-range exchange interaction in the case of the mean-field or a short-range interaction in the case of the Heisenberg or Ising models [1].

$Ga_{1-x}Mn_xAs$ semiconductors have been studied intensely over the last decade and have become a model system for diluted ferromagnetic semiconductors [2-4]. It is now widely accepted that the ferromagnetism in $Ga_{1-x}Mn_xAs$ arises from the hole mediated exchange interaction between the local magnetic moments of the Mn and it is well described by mean-field-like Zener model [3]. However, very recently, the mean-field-like behavior of $Ga_{1-x}Mn_xAs$ was questioned [5], since the critical exponent of the magnetization for the samples with effective Mn concentrations of 8% and 10% reveal the value of $\beta = 0.407(5)$ consistent with the short-range Heisenberg model for disordered magnetic system. From this result, authors made conclusions that the classical approach of the indirect exchange mechanism by free carriers for ferromagnetic ordering in $Ga_{1-x}Mn_xAs$ with high concentrations of Mn is questionable. Also, recently, the Curie point singularity in the temperature derivative of resistivity in $Ga_{1-x}Mn_xAs$ with nominal Mn concentration ranging from 4.5 % to 12.5 % has been investigated [6]. Using the similarity between the critical behaviors of the temperature derivative of the resistivity $d\rho/dt$, where $t = |T/T_c -1|$ is the reduced temperature, and the specific heat for metallic ferromagnets [7] the critical exponent $\alpha$ of the specific heat has been estimated from the $\log(d\rho/dt)$ vs $\log(t)$ plots. All data sets collapsed into the common temperature dependence for $T < T_C$, as well as another common dependence was observed for $T > T_C$. However, no clear power-law behavior in $d\rho/dt$ on the either side of the transition has been observed. At the same time, the temperature dependence of the magnetization for the sample with $T_C = 185$ K reveals a power-law dependence with an approximate exponent $\beta = 0.3$–$0.4$, consistent with either Heisenberg or Ising behavior. On the other hand, very recently, the results on the critical behavior study of $Ga_{1-x}Mn_xAs$ by the specific heat have been published [8]. The value of the specific heat critical exponent $\alpha$ for the $Ga_{1-x}Mn_xAs$ with the Mn concentrations of 1.6% is close to the critical exponent $\alpha \approx 0.11$ for the three-dimensional (3D) Ising model. While, for the $Ga_{1-x}Mn_xAs$ with the Mn concentration of 2.6% the critical behavior is well described by the mean-field including 3D Gaussian fluctuations model. The critical behavior of a system exhibiting a second order phase transition is strongly affected by the range of interactions. In the limit of infinite interactions, the system is characterized by the mean-field scaling behavior. However, according to the well-known Ginzburg criterion [9], the mean-field-like behavior occurs even for finite interaction ranges, sufficiently far away from the critical temperature.

In this paper, we present the results of experimental study of the critical behavior of $Ga_{1-x}Mn_xAs$ with different concentrations of Mn (from 2% to 10%) by using thermal diffusivity measurements. Taking into account the relationship between specific heat C and thermal diffusivity D through the equation $C = K/\rho D$ (where $\rho$ is the density and K is the thermal conductivity), the inverse of the thermal diffusivity has the same critical behavior as the specific heat [10].

The $Ga_{1-x}Mn_xAs$ layers with different Mn concentrations were grown on semi-insulating (001) GaAs substrates by using MBE. The epilayers with thickness about 60 nm were grown at low temperature of 270 $^0$C with different temperatures of the Mn source. No post-growth thermal annealing was performed. The Mn concentration in the layers was estimated from X-ray diffraction measurements and it was additionally confirmed by X-ray microanalysis. The resistivity measurement was conducted by using LakeShore system equipped with a closed cycle cooling cryostat. The temperature dependence of magnetization was measured by a superconducting quantum interference device (SQUID) magnetometer. The thermal diffusivity was measured by using a photothermal method described elsewhere [11].

Figure 1 shows the temperature dependence of the magnetization for $Ga_{1-x}Mn_xAs$ samples with three different Mn concentrations. The measurements were conducted at the magnetic field of 10 Oe applied parallel to the sample plane after cooling of samples in a zero magnetic field. From magnetization curves the Curie temperatures of these samples have been determined. The Curie temperatures for samples A, B, and C are about 64 K, 73K, and 80K, respectively. The temperature dependence of the resisitivity for these samples, measured in a zero magnetic field, is shown in the inset of Fig.1. It is seen that the sample A with the Mn concentration of 2% demonstrates an insulating behavior, while the samples B and C with the Mn concentration of 3% and 6%, respectively show a metallic behavior. All samples showed a maximum (a rounded cusp) near Curie temperature $T_C$, marked by solid arrows. It should be noted that the resistive maxima coincide well with the respective Curie temperatures, determined from magnetization curves, within the experimental errors of temperature measurements.

Figure 2 shows the temperature dependence of the thermal diffusivity for samples A, B, and C and the thermal diffusivity of the GaAs substrate is shown by dashed line. All samples demonstrate a pronounced λ shaped peak, which indicates a second-order phase transition in these samples. The thermal diffusivity peak located close to the Curie temperature and therefore, it is attributed to the ferromagnetic-paramagnetic phase transition. As seen in Fig. 2, with an increase of the manganese concentration the thermal diffusivity peak increases in amplitude and shifts to higher temperatures. The thermal diffusivity peak for $Ga_{1-x}Mn_xAs$ samples A, B, and C was observed at 63.85 K, 72.59 K, and 79.91 K, respectively. In order to study the critical behavior of the magnetic phase transition in the proximity of the Curie temperature we analyzed the inverse of the thermal diffusivity with subtracting off a smooth background as a nonmagnetic contribution of the GaAs substrate and $Ga_{1-x}Mn_xAs$ layer. The nonmagnetic contribution of the $Ga_{1-x}Mn_xAs$ layers to the thermal diffusivity is supposed to be very close to the thermal diffusivity of the GaAs because the Mn concentrations in the samples investigated are relatively low. Figure 3a and 3b show the plots of the normalized to the peak maximum of the inverse of the thermal diffusivity versus the reduced temperature for the $Ga_{1-x}Mn_xAs$ samples at $T > T_C$ and $T < T_C$, respectively. The expected theoretical value of the specific heat critical exponent $\alpha = 0.11$ for the 3D Ising and $\alpha = 0.5$ for the mean-field including 3D Gaussian fluctuations models are shown by the solid and dashed lines, respectively. The contribution of Gaussian fluctuations to the specific heat is given by $\Delta C = C^{\pm} t^{-\alpha}$, where $\alpha = 2 - d/2$, and d is the dimensionality [12,13]. It is seen from Figs. 3a and 3b that the critical exponent $\alpha$ of the samples investigated clearly shows the crossover from 3D mean-field-like to 3D Ising-like value with decreasing of the temperature distance from the critical point. Figure 3c shows the normalized specific heat experimental data from Figs. 3 and 4 of Ref. 8 for the $Ga_{1-x}Mn_xAs$ samples with the Mn concentrations of 1.6% and 2.6% versus the reduced temperature, where the data for $T < T_C$ and $T > T_C$ are shown by the filled and empty circles, respectively. It is seen that the critical exponent $\alpha$ for the sample with 1.6% of Mn shows the crossover from 3D Ising-like to 3D mean-field-like behavior, whereas for the sample with 2.6% of Mn the value of critical exponent $\alpha$ does not show any Ising-like behavior up to $t = 0.001$, but it shows the crossover from 3D to 2D mean-field-like behavior with increasing of the temperature distance from the critical point. The Ising-like critical behavior observed for the $Ga_{1-x}Mn_xAs$ samples demonstrates an existence of a strong uniaxial magnetic anisotropy in $Ga_{1-x}Mn_xAs$ epitaxial layer, which has already been theoretically and experimentally demonstrated [14,15].

The crossover of the critical behavior is ruled by the parameter $t/G$, where $t$ is the reduced temperature and $G$ is the so-called Ginzburg number [16]. The non-classical critical behavior occurs for $t/G \ll 1$ and classical critical behavior is expected for $t/G \gg 1$. The value of the Ginzburg number for the samples investigated has been determined from Fig. 3 as a midpoint of the crossover from the mean-field to Ising critical behavior. The value of the Ginzburg number can be also obtained from the crossover function obtained numerically by using the renormalization-group theory [17], however, it was beyond the scope of this paper. As seen from Fig. 3, the Ginzburg number $G^-$ for $T < T_C$ is always larger than $G^+$ for $T > T_C$ for the samples investigated. Although the very procedure for obtaining G cannot provide the high accuracy and the differences between $G^+$ and $G^-$ are not so large, a systematic character of this phenomenon deserves consideration.

The value of Ginzburg number depends on the effective interactions length R as $G = G_0 R^{-2d/4-d}$ [16,18], where d is the dimensionality and R is given in units of the lattice constant and hence dimensionless. As $R \to \infty$, a mean-field-like behavior is observed, while close enough to the critical temperature $T_C$ for finite R, ones always observes the non-classical critical exponents, differing from those of the Landau theory [19]. The exchange interaction length can be calculated from the expression given in Ref. 19 which, for the three dimensional case, reduces to $G = 27/(\pi^4 R^6) \approx 0.27718 / R^6$. Therefore, from Figs. 3a and 3b for sample A we have obtained $R^+ \approx 2$, which is about 11.3 Å and $R^- \approx 1.75 \approx 9.8$ Å. For the $Ga_{1-x}Mn_xAs$ sample with 2.6% of Mn (Fig. 3c), the both of the Ginzburg numbers below and above $T_C$ are lower than 0.001 and therefore, the exchange interaction length for this sample $R^\pm > 14.4$ Å. In Ref. 5, the interaction length was estimated as the mean-free pass length $\lambda$ of holes and for the hole concentration of $p \approx 5 \times 10^{20}$ cm$^{-3}$ the $\lambda$ is on the order of 5Å. Such a small value makes the classical approach of the indirect exchange mechanism by free carriers doubtful. However, the exchange interaction between magnetic ions in $Ga_{1-x}Mn_xAs$ is mediated by holes which are spin oriented and therefore, the interaction range should be considered as a spin diffusion length rather than the mean-free path length of holes. There are several main mechanisms for spin relaxation of carriers in semiconductors: the Elliot-Yafet, D'yakonov-Perel' and Bir-Aronov-Pikus [20]. In diluted magnetic semiconductors the spin-flip scattering by the sp-d exchange interaction should be added and the sp-d exchange interaction is believed to be dominated spin relaxation mechanism in magnetic semiconductors [20]. Also, the spin-flip scattering by the interstitial Mn ions which are antiferromagnetically coupled with the substituted Mn acceptors should be taking into account. Hence, the spin relaxation time of holes in $Ga_{1-x}Mn_xAs$ might be very short, however, it can be still longer comparing with the momentum relaxation time. Figure 4 shows the temperature dependence of the thermal diffusivity for the $Ga_{1-x}Mn_xAs$ sample with the Mn concentration of 10%, which was annealed at 160 $^0$C for 24 hours in air [21]. The inset shows the temperature dependence of magnetization for this annealed sample. From Fig. 4 the Curie temperature about 124 K was determined. Figure 5 shows the plots of the normalized inverse of the thermal diffusivity versus the reduced temperature above and below the critical temperature. It is seen that this sample demonstrate the mean-field-like behavior up to t = $10^{-3}$. This result demonstrates that the low-temperature annealing of $Ga_{1-x}Mn_xAs$ increases the exchange interaction length. Therefore, the short-range exchange interaction observed in $Ga_{1-x}Mn_xAs$ samples is not intrinsic feature of $Ga_{1-x}Mn_xAs$, as it was supposed in Ref. 5, but it depends strongly on the quality of samples. The large value of the Ginzburg number $t_g$ = 0.35 observed in Ref. 5 for the $Ga_{1-x}Mn_xAs$ samples with the effective Mn concentrations of x = 0.08 - 0.10, shows that the exchange interaction in these samples is indeed of a short-range, however, our results demonstrate that the exchange interaction length can be sufficiently increased by choosing the proper growth and thermal annealing conditions.

In conclusion, we have experimentally studied the critical behavior of $Ga_{1-x}Mn_xAs$ near the Curie temperature by using the thermal diffusivity measurements. For the most part of as-grown samples, the crossover from the mean-field-like to the Ising-like critical behavior of the specific heat critical exponent α, with approaching close to the Curie temperature, has been observed. This result demonstrates that the exchange interaction length between magnetic ions in $Ga_{1-x}Mn_xAs$ is finite. However, the as-grown sample with a moderate concentration of Mn (2.6%), as well as the sample with a high concentration of Mn (10%) after the low temperature annealing, demonstrate the mean-field-like behavior up to $t = 10^{-3}$. These results demonstrate that the exchange interaction length in $Ga_{1-x}Mn_xAs$ might be much larger than of 5Å determined in Ref.5.

We acknowledge gratefully Vit Novák and Karel Výborný for the annealed samples, and Tomasz Dietl for helpful discussions. This work was supported by the National Research Foundation (NRF) of Korea grant funded by the Ministry of Education, Science and Technology (MEST, No. 2011-0000016).

Figure captions

Fig. 1. Temperature dependence of the magnetization for the $Ga_{1-x}Mn_xAs$ with (a) 2%, (b) 3%, and (c) 6% of Mn measured in a magnetic field of 10 Oe. Inset: Temperature dependence of the resistivity for the same of $Ga_{1-x}Mn_xAs$ samples measured at a zero magnetic field.

Fig. 2. Temperature dependence of the thermal diffusivity for the $Ga_{1-x}Mn_xAs$ with (a) 2%, (b) 3%, and (c) 6% of Mn. The thermal diffusivity of the GaAs substrate is shown by dashed line.

Fig. 3. The normalized inverse of the thermal diffusivity versus the reduced temperature for the $Ga_{1-x}Mn_xAs$ samples with 2% (red circles), 3% (green circles), and 6% (blue circles) of Mn for (a) $T > T_C$ and (b) $T < T_C$, respectively. The theoretically expected dependencies for the 3D Ising and the mean-field with 3D Gaussian fluctuations models, are shown by the solid and dashed lines, respectively. (c) The normalized data of the magnetic specific heat for the $Ga_{1-x}Mn_xAs$ samples with 1.6% (red circles) and 2.6% (blue circles) of Mn, derived from the experimental data of Ref. 8. The filled and open circles represent the data for $T < T_C$ and $T > T_C$, respectively. The dashed-dotted line $\alpha = 1$ represents the theoretical dependence of the specific heat for the mean-field model with 2 D Gaussian fluctuations.

Fig. 4. Temperature dependence of the thermal diffusivity for the $Ga_{1-x}Mn_xAs$ with 10% of Mn, annealed at 160 $^0$C for 24 hours in air. The thermal diffusivity of the GaAs substrate is shown by dashed line. Inset: Temperature dependence of the magnetization for the annealed sample.

Fig. 5. The normalized inverse of the thermal diffusivity versus the reduced temperature for the annealed sample. The filled and open circles represent the data for $T < T_C$ and $T > T_C$, respectively.

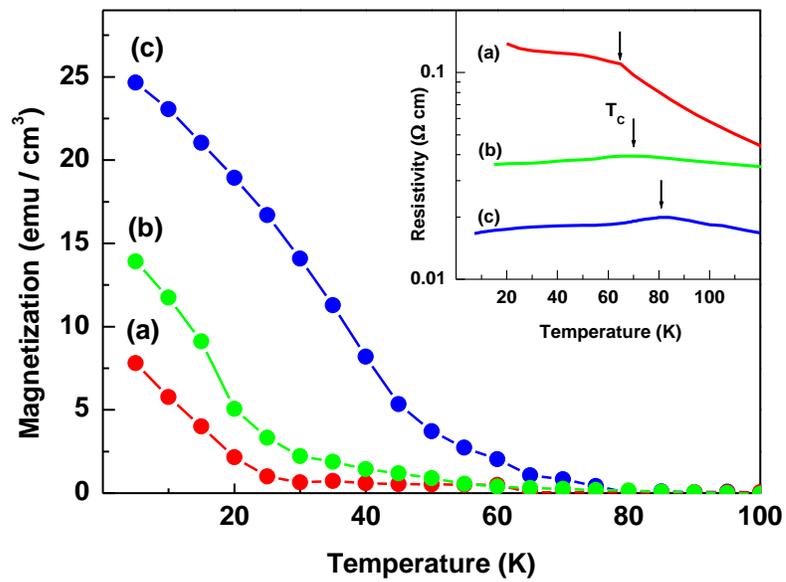

Figure 1

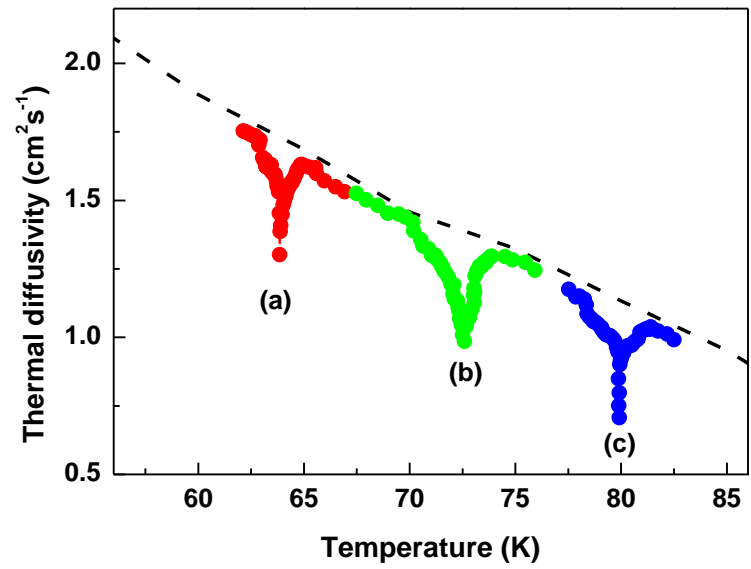

Figure 2

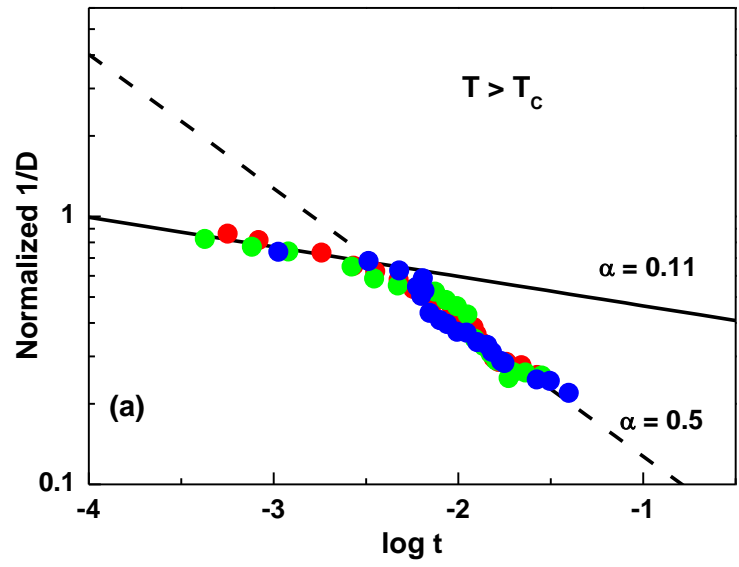

Figure 3a

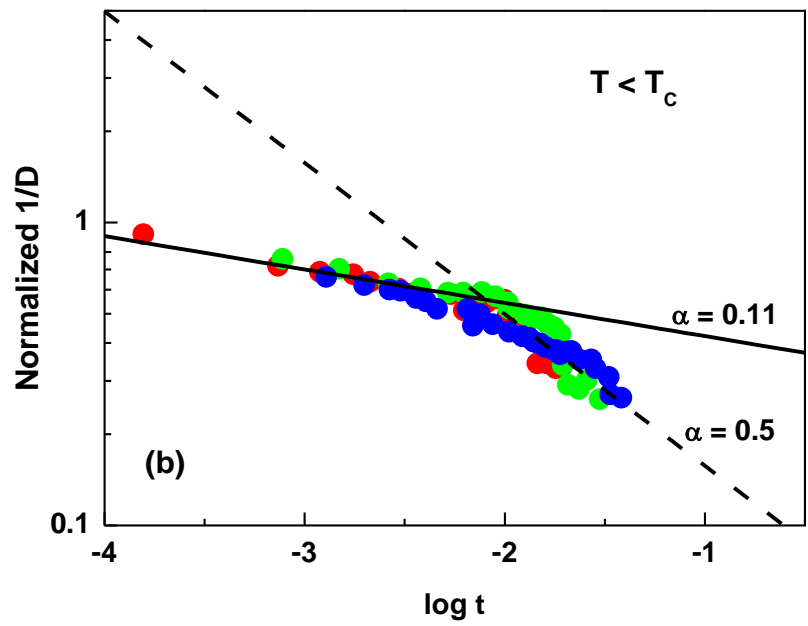

Figure 3b

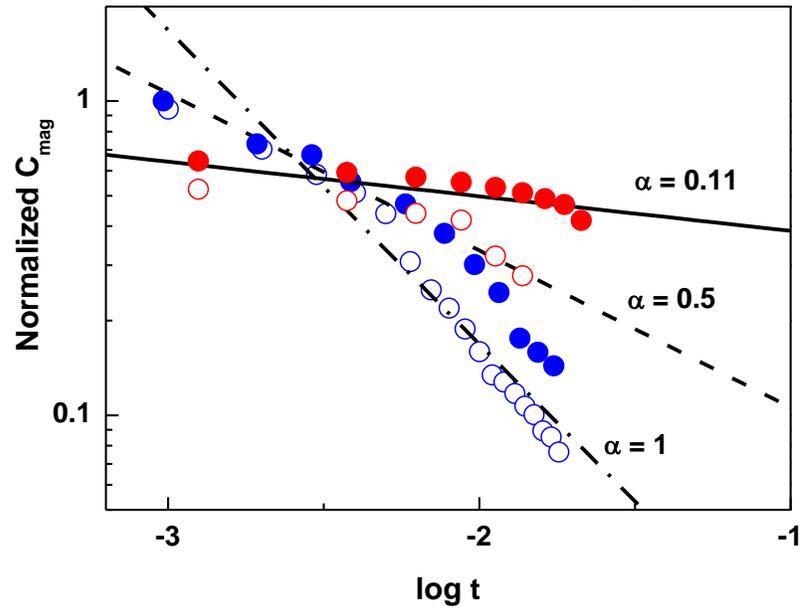

Figure 3c

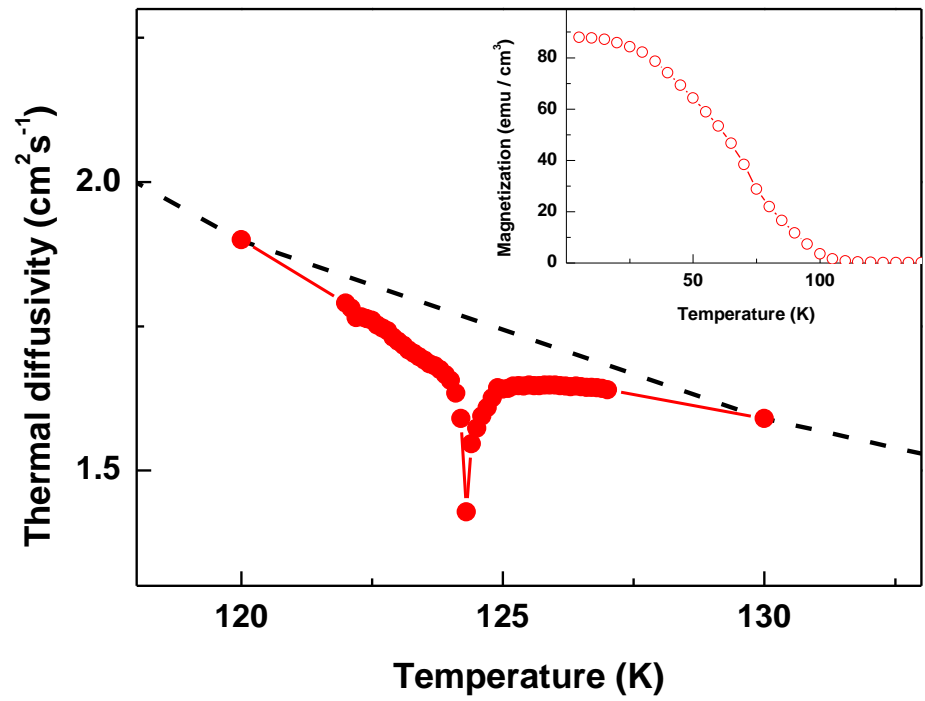

Figure 4

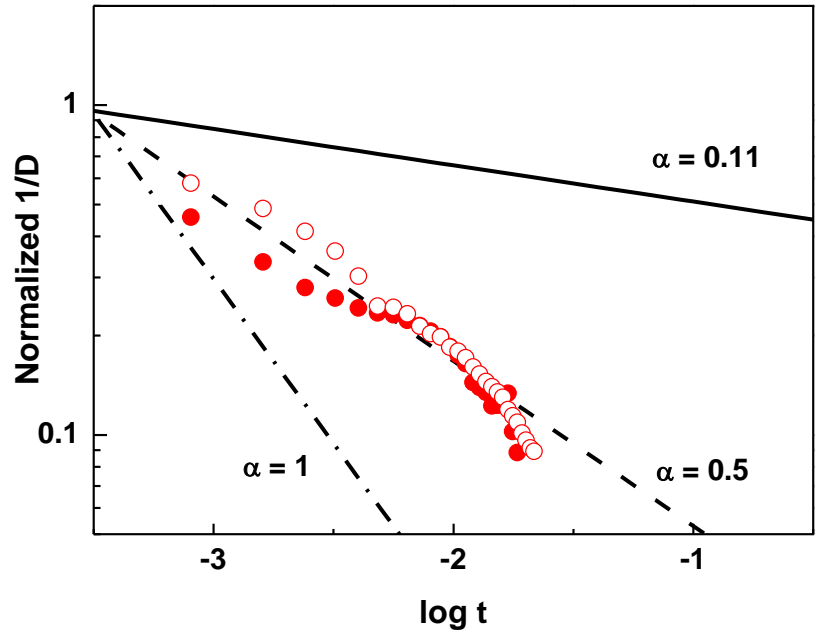

Figure 5